\documentclass[amssymb, nobibnotes, aps,pra,preprint]{revtex4-1}

\usepackage{graphicx,epsfig,epstopdf}
\usepackage{amsmath} 
\usepackage{bm}
\usepackage{color}
\begin{document}

\title{Nonlinear fluctuations and dissipation in matter revealed by quantum light}


\author{Shaul Mukamel}
\email{smukamel@uci.edu}

\author{Konstantin E. Dorfman}
\email{kdorfman@uci.edu}


\affiliation{Department of Chemistry, University of California, Irvine,
California 92697-2025, USA}
\date{\today}
\pacs{42.50 ct, 42.50 DV, 42.50 Lc, 33.80 b}

\begin{abstract}
Quantum optical fields offer numerous control knobs which are not available with classical light and may be used for monitoring the properties of matter by novel types of spectroscopy. It has been recently argued that such quantum spectroscopy signals can be obtained by a simple averaging of their classical spectroscopy counterparts over the Glauber-Sudarshan quasiprobability distribution of the quantum field; the quantum light thus merely provides a novel gating window for the classical response functions.  We show that this argument only applies to the linear response and breaks down in the nonlinear regime. The quantum response carries additional valuable information about response and spontaneous fluctuations of matter that may not be retrieved from the classical response by simple data processing. This is connected to the lack of a nonlinear fluctuation-dissipation relation.
\end{abstract}

\maketitle


\vspace{0.5cm}

\section{Introduction}

Quantum optical fields offer many types of unique control knobs (parameters of the photon wavefunction) that may be used to simplify, manipulate and display spectroscopic signals. The quantum nature of light is widely used for quantum computing and information processing where the key goal is the manipulation of complex light fields by simple matter systems (qubits) \cite{hor09,pan12,gis02,boy06,gio11,kal14,lvo09,shi03}. Common spectroscopic applications on the other hand use classical light in order to learn about matter by varying pulse frequencies, delays and polarizations. Spectroscopy with quantum light, known as quantum spectroscopy \cite{sal98,ros09,sch13,lee06,day05}, is made possible by recent progress in photon quantum state engineering \cite{pen91,bel10,bra99,zav04,zah08,afe10,kie11,sil14}. Quantum spectroscopy had been applied to overcome the time/frequency Fourier uncertainty in Raman signals \cite{dor140} to control two-exciton states in photosynthetic complexes \cite{sch13} and to obtain nonlinear signals with weak fields, thanks to the improved scaling of signals with light intensity: e.g. two photon absorption with entangled photons scales linearly rather than quadratically with pump intensity \cite{sal98,ros09}.  The classical response functions (CRF), which describe the response of a quantum system to classical fields, are causal; the field affects the   material system but the system does not affect the field. The situation is fundamentally different when two quantum systems (matter and field in our case) interact. Now the response and spontaneous fluctuations of both systems mix and causality does not apply \cite{coh03}.  Quantum signals thus carry matter information other than the CRF, and consequently quantum nonlinear spectroscopy signals may not be retrieved merely by data processing of classical signals. This is good news making the quantum response much more exciting; quantum light reveals new types of information and phenomena related to the interplay of response and fluctuations, which is not accessible by classical light \cite{ros10}.

In a series of publications on quantum spectroscopy in semiconductors \cite{kir11,alm14,moo14} it has been argued that the underlying matter information revealed by quantum fields is the same as in the classical field case.  The argument starts with the Glauber Sudarshan P representation which expresses the field density matrix as an integral over coherent state density matrices $|\beta\rangle\langle\beta|$ weighted by a quasi-probability distribution $P(\beta)$
\begin{align}\label{eq:Pb}
\hat{\rho}=\int d^2\beta P(\beta)|\beta\rangle\langle\beta|.
\end{align}
A quasiprobability distribution is a representation of the density matrix that allows to recast observables as a classical-looking average over that distribution \cite{gla07}. However this function is not a genuine probability distribution since it can have both positive and negative values. Such distributions are common in quantum optics, most notably in the theory of the laser \cite{scu97}.
It has been suggested \cite{kir11,alm14,moo14} that since the response of a material system to a field initially prepared in a coherent state $|\beta\rangle$ is given by the classical response function CRF, the quantum response  $R_{QM}$ may be recast as an average of the classical response $R_{|\beta\rangle}$ with respect to this quasi-probability  
\begin{align}\label{eq:Rqm}
R_{QM}=\int d^2\beta P(\beta)R_{|\beta\rangle}.
\end{align}
The thrust of this representation is that the quantum field merely provides a novel gating window for the classical response function (CRF); a complete knowledge of the CRF is enough to compute the response to any quantum field, and the quantum response function (QRF) may be then recovered from the classical response function (CRF) by simple data processing. If correct, this makes quantum spectroscopy less interesting since it does not carry fundamentally new matter information.

Here we show that Eq (\ref{eq:Rqm}) only holds for the linear response, and does not apply to the nonlinear response. In Section II we illustrate the additional information about quantum paths provided by the quantum response, which is missed by classical fields by an example calculation of third order nonlinear response to the quantum field. We then use superoperators in Section III to connect this more broadly to the absence of a nonlinear fluctuation dissipation theorem: spontaneous fluctuations and response are only uniquely related in the linear regime \cite{has91}, but not when they are nonlinear. Some nonlinear fluctuation-dissipation relations have been proposed for specific models under limited conditions \cite{lip08,boc81,ber01} but there is no universal relation of this type \cite{kry12}.

\section{Third order nonlinear response to the quantum field}

\begin{figure}[h]
 \centering
 \includegraphics[width=0.75\textwidth]{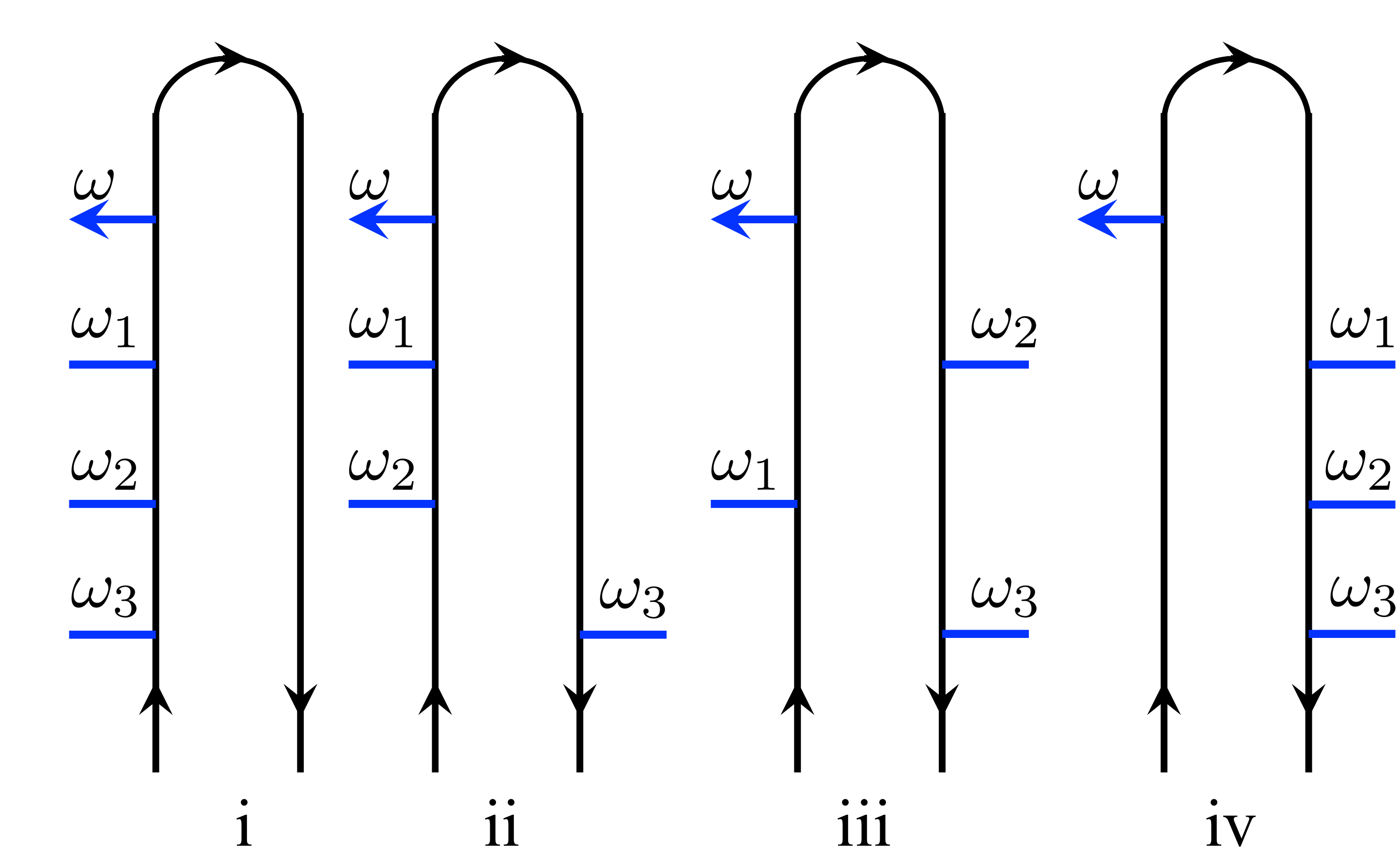}
\caption{(Color online) Loop diagrams for the third order nonlinear response. The time runs along the loop forward in left branch and backward in the right branch. Time translation invariance enforces $\omega-\omega_1-\omega_2-\omega_3=0$. For diagram rules see \cite{Rah10}. Since we do not invoke rotating wave approximation, field-matter interactions depicted by blue lines do not have arrows indicating whether it is creation or annihilation operator. Interaction with $\omega$ has an arrow as it represents the signal in the form of Eq. (\ref{eq:Sdef}) where having negative frequency component is necessary.}
 \label{fig:Pab}
\end{figure}

We start by a simple example that illustrates why Eq. (\ref{eq:Rqm}) fails. Consider a multilevel quantum system that interacts with a quantum optical field $E(t)$ via the dipole operator
\begin{align}
H(t)=E(t)V(t),
\end{align}
where $E(t)=\tilde{E}(t)+\tilde{E}^{\dagger}(t)$ is the electric field operator that annihilate (fist term) and create (second term) a photon. The dipole operator $V(t)=\tilde{V}(t)+\tilde{V}^{\dagger}(t)$ similarly contains lowering and raising operators.  The frequency dispersed transmission of the field is given by a rate of change of photon number which can be recast as
\begin{align}\label{eq:Sdef}
S(\omega)=\mathcal{I}\langle \tilde{E}^{\dagger}(\omega)P(\omega)\rangle_f,
\end{align}
where $\mathcal{I}$ denotes imaginary part, $\langle ...\rangle=\text{Tr}[...\rho_f(t)]$ is the trace over the quantum field degrees of freedom in the space of quantum field that is created by field-matter interaction and $P(\omega)=\int due^{i\omega t}P(t)$ is a Fourier transform of the polarization operator. The third order nonlinear response of the system is given by four loop diagrams shown in Fig. \ref{fig:Pab} (for rules see \cite{Rah10}) and can be read as
\begin{align}
S_{QM}(\omega)=S_i(\omega)+S_{ii}(\omega)+S_{iii}(\omega)+S_{iv}(\omega),
\end{align}
where
\begin{align}\label{eq:Si}
S_i(\omega)=\mathcal{I}\int\frac{d\omega_1}{2\pi}\int\frac{d\omega_2}{2\pi}\int\frac{d\omega_3}{2\pi}\langle \tilde{E}^{\dagger}(\omega)E(\omega_1)E(\omega_2)E(\omega_3)\rangle_f F_i(-\omega;\omega_1,\omega_2,\omega_3),
\end{align}
\begin{align}
S_{ii}(\omega)=\mathcal{I}\int\frac{d\omega_1}{2\pi}\int\frac{d\omega_2}{2\pi}\int\frac{d\omega_3}{2\pi}\langle E(\omega_3) \tilde{E}^{\dagger}(\omega)E(\omega_1)E(\omega_2)\rangle_f F_{ii}(-\omega;\omega_1,\omega_2,\omega_3),
\end{align}
\begin{align}
S_{iii}(\omega)=\mathcal{I}\int\frac{d\omega_1}{2\pi}\int\frac{d\omega_2}{2\pi}\int\frac{d\omega_3}{2\pi}\langle E(\omega_3)E(\omega_2)\tilde{E}^{\dagger}(\omega)E(\omega_1)\rangle_f F_{iii}(-\omega;\omega_1,\omega_2,\omega_3),
\end{align}
\begin{align}\label{eq:Siv}
S_{iv}(\omega)=\mathcal{I}\int\frac{d\omega_1}{2\pi}\int\frac{d\omega_2}{2\pi}\int\frac{d\omega_3}{2\pi}\langle E(\omega_3)E(\omega_2)E(\omega_1)\tilde{E}^{\dagger}(\omega)\rangle_f F_{iv}(-\omega;\omega_1,\omega_2,\omega_3),
\end{align}
where matter pathways are given by
\begin{align}\label{eq:Fi}
F_i(-\omega;\omega_1,\omega_2,\omega_3)=\langle VG(\omega)VG(\omega-\omega_1)VG(\omega_3)V\rangle2\pi\delta(\omega_1+\omega_2+\omega_3-\omega),
\end{align}
\begin{align}
F_{ii}(-\omega;\omega_1,\omega_2,\omega_3)=\langle VG^{\dagger}(\omega_3)VG(\omega-\omega_3)VG(\omega_2)V\rangle2\pi\delta(\omega_1+\omega_2+\omega_3-\omega),
\end{align}
\begin{align}
F_{iii}(-\omega;\omega_1,\omega_2,\omega_3)=\langle VG^{\dagger}(\omega_3)VG^{\dagger}(\omega-\omega_1)VG(\omega_1)V\rangle2\pi\delta(\omega_1+\omega_2+\omega_3-\omega),
\end{align}
\begin{align}\label{eq:Fiv}
F_{iv}(-\omega;\omega_1,\omega_2,\omega_3)=\langle VG^{\dagger}(\omega_3)VG^{\dagger}(\omega-\omega_1)VG^{\dagger}(\omega)V\rangle2\pi\delta(\omega_1+\omega_2+\omega_3-\omega)
\end{align}
and $G(\omega)=1/\hbar[\omega-H_0/\hbar+i\epsilon]$ is the retarded Hilbert space Green's function that governs the forward time propagation whereas $G^{\dagger}(\omega)=1/\hbar[\omega-H_0/\hbar-i\epsilon]$ is the corresponding advanced Green's function that governs the backward time propagation.

 For classical fields one can replace field operators $E$ by their expectation values $\mathcal{E}=\langle E\rangle$. In this case the four field correlation functions in Eqs.  (\ref{eq:Si}) - (\ref{eq:Siv}) are the same and the total signal  is given by
\begin{align}\label{eq:Sc}
S_{\text{cl}}(\omega)=\mathcal{I}\int\frac{d\omega_1}{2\pi}\int\frac{d\omega_2}{2\pi}\int\frac{d\omega_3}{2\pi}\tilde{\mathcal{E}}^{*}(\omega)\mathcal{E}(\omega_1)\mathcal{E}(\omega_2)\mathcal{E}(\omega_3)\chi^{(3)}(-\omega;\omega_1,\omega_2,\omega_3),
\end{align}
where $\chi^{(3)}(-\omega;\omega_1,\omega_2,\omega_3)=\sum_j F_j(-\omega;\omega_1,\omega_2,\omega_3)$ is the third order nonlinear susceptibility that represents the response of the system to classical fields.
 
The key difference between Eqs.  (\ref{eq:Si}) - (\ref{eq:Siv}) and Eq. (\ref{eq:Sc}) is that in the quantum response each of the four pathways (Eqs. (\ref{eq:Fi}) - (\ref{eq:Fiv})) is gated by a different field correlation function whereas in the classical response the gates are identical allowing to combine the four diagrams into a single classical response function ($\chi^{(3)}$ in this case). The four gates differ by the position of the detected field $\tilde{E}^{\dagger}(\omega)$ along the loop (fourth, third, second and first along the loop for diagrams $i$, $ii$, $iii$, and $iv$, respectively).

 
To explain the above result we note that a classical (coherent) state of the field does not change in the course of field-matter interactions. It is therefore independent of the ordering between field operators. The matter dynamics is then decoupled from the field and can be studied separately by CRF.  In the case of quantum field  the state of the field does change in the course of the process, as is evident from the fact that different diagrams contain different correlation functions of the field operators. We must therefore  do the calculation in the joint field-matter space, whereas in the classical case the field factorizes out since the state of the classical field is unchanged. By working in the joint field plus matter space one  keeps track of the both matter and the field. 


We can alternatively explain the difference as follows. In general, the quantum nature of the field enters the QRF in two ways (i) through the initial quantum distribution of the field, and  (ii) through the fact that both the field and the matter states vary during the course of   their coupled evolution that generates the response. Field and matter become entangled via a path integral in their joint space.  Eq. (\ref{eq:Rqm}) only accounts for (i) but ignores (ii). The response to a field initially prepared in a coherent state $|\beta\rangle$ coincides with the classical response only if all the field operators involved in the specific signal are normally ordered.  A coherent field state then remains unaltered in the course of the evolution and only point (i) applies.  However this is not generally the case. The CRF totally misses point (ii) and consequently does not carry all the information about field/matter entanglement that enters into the QRF. 

The experiments reported in \cite{kir11,alm14,moo14} are nonlinear pump probe with classical light. They do not involve quantum light so that the experimental data are fine. The classical signals were then expanded in the form of Eq. (\ref{eq:Rqm}) using various $P(\beta)$ and it has been argued that the resulting $R_{QM}$  is the QRF as explained above. This approach only takes into account point (i) but not (ii). The initial state of the field can be always represented by Eq. (\ref{eq:Pb}). However it does not take into account the entanglement of matter and field that affects the quantum response as shown in Eqs. (\ref{eq:Si}) - (\ref{eq:Siv}). The claim that by decomposing the classical signals using the Glauber-Sudarshan distribution it is possible to extract the quantum response is an unjustified conjecture that has not been tested by the above experiments.

\section{Connection to nonlinear fluctuation-dissipation relations}

We now use superoperator notation \cite{ros09,ros10} to show more broadly why the quantum response is different from the classical one so that Eq. (\ref{eq:Rqm}) is violated.  With each quantum operator $V$  we associate two superoperators   $V_+$ (anti-commutator) and  $V_-$  (commutator) defined by their action on another operator  $X$
 \begin{align}
V_+X=\frac{1}{2}(VX+XV),\quad V_-X=VX-XV.
\end{align}

For the spectroscopy applications considered here $V$ is the dipole operator. The interacting Hamiltonian superoperator is given by
\begin{align}
H_{int-}=E_+V_-+E_-V_+
\end{align}
We now consider a  system of two noninteracting atoms $1$ and $2$. The response of the subsystem, e.g. atom $1$  can be calculated by taking an expectation value of operator $O_1$
\begin{align}
\text{tr}[O_1\rho(t)]=\text{tr}&\left[O_1\mathcal{T}\exp\left(-\frac{i}{\hbar}\int^t[E_+(t')V_{1-}(t')+E_-(t')V_{1+}(t')]dt'\right)\right.\notag\\
&\left.\times\exp\left(-\frac{i}{\hbar}\int^t[E_+(t')V_{2-}(t')+E_-(t')V_{2+}(t')]dt'\right)\rho_{10}\rho_{20}\rho_{ph0}\right]
\end{align}
where we factorized the initial density operator into a product of parts corresponding to atom $1$, atom $2$, and field.
For a classical field $E_-=0$ (the commutator vanishes) and therefore there are only $V_{2-}$ operators for atom $2$. Any order correlation function of atom $2$ would be given in the form of $\langle V_{2-}V_{2-}...V_{2-}\rangle=0$, since the trace of commutator is zero. However, for a quantum field the correlation function of the field will involve $E_-$ and therefore the matter correlation function of atom $2$ involves $V_{2+}$ and does not vanish.

The time evolution of two coupled quantum systems and the field is generally given by a sum over Feynman paths in their joint phase space. Order by order in the coupling, dynamical observables can be factorized into products of correlation functions defined in the individual spaces of the subsystems. These correlation functions represent both {\it causal} response and {\it non-causal} spontaneous fluctuations \cite{coh03,ros10}. 

$\langle V_+V_+\rangle$  and $\langle V_+V_-\rangle$ are the only two quantities that contribute to the linear response   ($\langle V_-V_-\rangle$ vanishes since it is the trace of a commutator). However, the two are not independent since they are related by the universal fluctuation -dissipation relation \cite{has91}.  
\begin{align}\label{eq:fdr}
C_{++}=\frac{1}{2}\coth(\beta\hbar\omega/2)C_{+-}(\omega).
\end{align}
Here
\begin{align}
C_{+-}(\omega)=\int d\tau\langle V_{+}(\tau)V_-(0)\rangle e^{i\omega\tau}
\end{align} 	
Is the response function whereas
\begin{align}
C_{++}(\omega)=\int d\tau\langle V_{+}(\tau)V_+(0)\rangle e^{i\omega\tau}
\end{align} 
 represents spontaneous fluctuations.					

The classical response function $C_{+-}(\omega)$  thus carries all relevant information about linear radiation matter coupling, including the quantum response. In the nonlinear regime the CRF is a specific causal combination of matter correlation functions given by one ``+'' and several ``-'' operators. e.g  $\langle V_+V_-V_-V_-\rangle$  for the third order response. However, the quantum response may also depend on the other combinations. To  $n^{\text{th}}$   order in the external field the CRF $\langle V_{+}(\omega_{n+1})V_{-}(\omega_n)...V_{-}(\omega_2)V_-(\omega_1)\rangle$  is one member of a larger family of  $2^n$ quantities $\langle V_{+}(\omega_{n+1})V_{\pm}(\omega_n)...V_{\pm}(\omega_2)V_\pm(\omega_1)\rangle$  representing various combinations of spontaneous fluctuations (represented by $V_+$ ) and impulsive excitations (represented by  $V_-$). For example, an "all +" quantity such as  $\langle V_+V_+V_+V_+\rangle$ represents purely spontaneous fluctuations. The CRF does not carry enough information to reproduce all  $2^n$  possible quantities which are accessible by quantum spectroscopy. The deep reason why the CRF and QRF are not simply related in is the lack of a fluctuation-dissipation relation in the nonlinear regime \cite{lip08,boc81,ber01,kry12}.

Note, that the example considered in section II is related to the fact that due to the change of the state of quantum field in the course of the optical process different components of the nonlinear response are multiplied by different detection windows governed by field correlation functions. This effect involves both $E_+$ as well as $E_-$, which appears if we try to reorder field operators in correlation functions, corresponding to different diagrams and superoperator algebra is another way to described quantum field effects in the clear way. The commutator of the field $E_-$ is intrinsically related to vacuum modes of the field which may induce coupling between noninteracting parts of the system. One example where such an effect arising from $E_-$ is combined with the appearance of collective resonances, which occurs for $E_+$ has been recently investigated in the context of harmonic systems \cite{gle15}. The response of classical or quantum harmonic oscillators coupled linearly to a classical field is strictly linear; all nonlinear response functions vanish identically. We have recently shown that quantum modes of the radiation field that mediate interactions between harmonic oscillator resulted in nonlinear susceptibilities. A third order nonlinear transmission of the optical field yields collective resonances that involve pairs of oscillators and are  missed by the conventional quantum master equation treatment \cite{Dor130}.

\section{Conclusions}

In summary, the relevant matter information in classical spectroscopy  may be recast in terms of nonlinear response functions $\langle V_+V_-...V_-\rangle$ (one ``+'' and several ``-'' operators). These represent $R_{|\beta\rangle}$in Eq. (\ref{eq:Rqm}). The classical response is causal (the field affects the system only at later times). Quantum spectroscopy signals carry qualitatively richer information that combines causal response with noncausal spontaneous fluctuations $\langle V_+...V_+V_-....V_-\rangle$ (several ``+'' and ``-'') and is missed by classical signals. The signals reflect the entanglement of field and matter in the course of their coupled evolution. Quantum spectroscopy signals may not be obtained by a simple data processing of their classical spectroscopy counterparts. It follows from the fluctuation dissipation theorem (Eq. (\ref{eq:fdr})) that linear quantum and classical spectroscopy signals carry the identical information. The new information in nonlinear quantum spectroscopy stems from the absence of such universal theorem in the nonlinear regime. Quantum spectroscopy experiments may be thus designed to help understand the interplay of response and fluctuations in many body systems. This qualitatively novel information is totally missed by the classical response.

\acknowledgements

This work was supported by the U.S. Department of Energy, Office of Science, Basic Energy Sciences under Award No. DE-FG02-04ER15571, and  the National Science Foundation (Grant No. CHE-1361516).




%

\end{document}